# Versatile Electrostatic Assembly of Nanoparticles & Polyelectrolytes: Coating, Clustering and Layer-by-Layer Processes

J.-P. Chapel[a] and J.-F. Berret[b]

[a] *Centre de Recherche Paul Pascal (CRPP), UPR CNRS 8641, Université Bordeaux 1, 33600 Pessac (France)*
[b] *Matière et Systèmes Complexes, UMR 7057 CNRS Université Denis Diderot Paris-VII, Bâtiment Condorcet, 10 rue Alice Domon et Léonie Duquet, 75205 Paris (France)*

Key words: inorganic nanoparticles, ion-containg polymers, electrostatic complexation, layer-by-layer, capsules,

**Abstract :** Engineered nanoparticles made from noble metals, rare-earth oxides or semiconductors are emerging as the central constituents of future nanotech developments. In this review, a survey of the complexing strategies between nanoparticles and oppositely charged polyelectrolytes developed during the last three years and based on electrostatic interactions is presented. These strategies include the one-step synthesis of stable and functionalized nanoparticles, the one- and multilayer coating of individual nano-objects, the controlled clustering of particles and the generation of capsules and thin films with superior functionalities. Among the formulation processes reported, three main classes are identified : the *direct mixing* route, the *desalting transition* pathway and the well-known *layer by layer* method. Finally, some latter developments, trends and applications of electrostatic assemblies in materials science and nanomedicine are highlighted.

# Outline

# Abbreviations

| | |
|---|---|
| CNT | carbon nanotube |
| DNA | deoxyribonucleic acid |
| El*b*L | electrostatic layer-by-layer |
| Gd-DOTA | gadoterate meglumine |
| L*b*L | layer-by-layer |
| MRI | magnetic resonance imaging |
| NP | nanoparticle |
| PAA | poly(acrylic acid) |
| PAH | poly(allylamine hydrochloride) |
| PDADMAC | poly(diallyldimethylammonium chloride) |
| PE | polyelectrolyte |
| PEG | poly(ethylene glycol) |
| PEI | poly(ethyleneimine) |
| PSS | poly(styrene sulfonate) |
| SEM | scanning electron microscopy |
| siRNA | small interfering ribonucleic acid |

# 1 – Introduction

Engineered nanoparticles (NPs) made from noble metals, rare-earth oxides or semiconductors are emerging as the central constituents of future nanotech developments. Interest stems from the combination of complementary attributes, such as a size in the nanometer range and unique physical features including high reactivity, electronic, magnetic or optical properties that can be exploited in a variety of applications ranging from catalysis, photovoltaics and coatings to nanomedicine [1-3]. During the last two decades, an extensive body of research has been devoted to the development of newly shaped NPs for materials science and nanomedicine. Nowadays, NPs are not only monodisperse spheres of few nanometers. They are also rods, triangles, disks, belts, ribbons, tetrapods, capsules, shells, needles etc… each of these morphologies bringing noticeable and unprecedented properties. More remarkably, nanomaterials are now synthesized from well-established and controlled chemistry, and for some of them like gold NPs and carbon nanotubes their production has reached significant levels.

Numerous synthesis routes thus exist, but one key limitation for their practical use is the restricted colloidal stability of the generated inorganic NPs in particular in water-borne solutions. Because of their high specific surface and chemical reactivity, as-synthesized NPs are sensitive to any change in physico-chemical environment such as pH, ionic strength, temperature or concentration, leading in some cases to their irreversible aggregation with possibly the loss of their physical size-related attributes [4,5]. In the quest of resilient coatings and coating strategies, neutral and/or charged polymers have become a paradigm. It is now recognized that in many situations such macromolecules outperform the classical low molecular weight ligands or oligomers used extensively in the past. NPs are usually coated with an organic layer providing solubility, long-term colloidal stability and functionalization. The reasons for the outstanding stabilization behavior of polymers are now well understood : *i)* polymers at interfaces form a brush of stretched chains which acts as a steric barrier against coagulation. Chosen to offset the van der Waals attraction, the range of the repulsion is monitored both by the molecular weight of the polymer and the grafting density. *ii)* When the polymer is bearing charges along its backbone, as in polyelectrolytes (PEs), the organic layer imparts an additional electrostatic repulsion between NPs. The combination of steric and electrostatic is often referred to *electrosteric* interaction. *iii)* The dangling ends of the tethered chains are chemical sites for further functionalization.

In the present review, a survey of the complexing strategies of NPs with oppositely charged PEs developed during the last 2 – 3 years and uniquely based on electrostatic interactions is presented. This approach takes advantage of the structural charges existing at the NPs surface, either through ionization of the crystallographic planes (or facets) in contact with a polar solvent, or through the adsorption of low molecular weight charged ligands like citric acid for example. For obvious reasons, the "grafting-from" and "grafting-to" methods which consist to polymerize monomers directly on the prefunctionalized NP surfaces [3], or to the covalent coupling or hydrogen bonding driven adsorption of functional macromolecules respectively will not be treated in this review.

Here, a strong emphasis will be put on the formulation strategies available in the literature. These strategies include the one-step synthesis of stable NPs in aqueous environments, the one- and multilayer coating of single NPs with PEs, the clustering or bulk complexation of oppositely charged NPs and PEs and the generation of NPs/PEs hybrid colloidal capsules or thin films for the design of multifunctional objects.

Electrostatic complexation between NPs and PEs has indeed numerous advantages. The technique is essentially a formulation-oriented process, performed under normal atmospheric conditions of pressure and temperature. As such, it displays much higher yield as compared to the "grafting-from" or "grafting-to" covalent reactions. Among the formulation processes reported so far, three main classes are identified : *i)* the *direct mixing* route, in which the NP and PE dispersions are simply mixed together e.g. by pouring one liquid into another, *ii)* the *desalting transition* pathway, which consists in screening the electrostatic interactions by addition of a large amount of salt, and subsequently in removing the excess progressively, e.g. by dialysis; *iii)* the well-known *layer by layer* (L*b*L) method which allows an alternate "decoration" of a given charged colloid or macroscopic surface with oppositely charged NPs and/or PEs

On a different aspect, formulation-based methods are interesting too since they allow scaling up any process to large quantities and workloads, as required by industrial constraints. Last but not least, the electrostatic complexation can be *fine-tuned* to co-assemble matter at the nanometer scale in a very elaborate manner. It goes well beyond the simple core-shell structures envisioned for coating NPs. Electrostatic complexation can be exploited to generate hybrid aggregates of controlled sizes between 100 nm and 100 μm, or L*b*L assemblies in the forms of capsules and thin films with advanced functionalities. In the last years, the versatility and the robustness of this simple technique were recognized, and attracted an increasing attention in the field of nanotechnology. The present review will survey the later developments in coating, clustering and L*b*L assembly of NPs using electrostatic complexation and highlight the actual trends in materials science and nanomedicine applications.

In the following, the most commonly used polyanions like poly(acrylic acid) (PAA) and poly(styrene sulfonate) (PSS), or polycations like poly(diallyldimethylammonium chloride) (PDADMAC), poly(ethyleneimine) (PEI) and poly(allylamine hydrochloride) (PAH) will be named after their acronym.

# 2 – Coating individual nanoparticles with polyelectrolytes

Polyelectrolytes are primarily used with inorganic NPs to provide colloidal stability to the dispersions. As mentioned in the introduction, this stability is a prerequisite for any water-borne applications. In the following, different types of recent formulations, involving charged homo- or copolymers electrostatically adsorbed as single or multilayer on the particles surface are reviewed

## 2.1 - Polyelectrolyte-assisted synthesis of nanoparticles

Most coating techniques are derived from the "grafting-to" approach. In few cases however, the PEs are introduced during the synthesis process, and serve as a nanoreactor or template. The generation of the NPs results then from the catalytic reduction of metallic salts in the presence of the PEs. The choice of the PEs is dictated by the charge of the metal ions, which need to be oppositely charged to facilitate the chain condensation. During the last 2 years, this technique was successfully applied to copper [6], palladium [7] and gold [8] NPs using PAA and carboxylate-based chains. Polycations such as PDADMAC, PEI and PAH were also utilized for the synthesis of platinum [9] or lead sulfide [10]. For the copper, platinum and lead sulfide, the PEs were assumed to form a stable adlayer with trains and loops electrostatically adsorbed at the surface of the particles, enhancing considerably their colloidal stability [6,9,10]. More specifically, Prucek *et al.* examined the synthesis of stable copper NPs assisted by poly(acrylate) chains [6]. These authors also showed that the controlled oxidation of the dispersions (obtained by removing the inert atmosphere above the liquid) resulted in the transformation of copper particles into $CuO_2$ nanocubes of average size 18 nm (Fig. 1a), both types of particles displaying excellent catalytic activity [6]. In the previously mentioned reports [6-10], the role of the PEs in the NPs formation was not fully elucidated.

## 2.2 - Adsorption of the first adlayer

Standard coating procedures of a single PE adlayer around sub-100 nm particles are generally based on the technique of L*b*L assembly that was designed in the first place for macroscopic surfaces [13]. Schneider and Decher recently described the formulation process in detail, using 13.5 nm citrate coated gold particles and PAH polycations as a model system [5]. The method is illustrated in Fig. 1b. To avoid flocculation, gold NPs were mixed to polymers away from the stoichiometry, *i.e.* with a large excess of cationic charges. This *direct mixing* procedure resulted in a spontaneous and maximum coverage of the particles *via* the complexation of the opposite charges. Optimum conditions were discussed as a function of the polymer concentration, polydispersity, charge ratio and ionic strength. This versatile technique allowed the addition of a second, a third etc… layer following the same deposition protocol (see Sect. 4.2). The L*b*L technique was implemented on cationic gold particles [14], and on gold nanorods [4,15] for applications in nanomedicine.

The main drawbacks of the LbL on NPs are an initial dilute mixing concentration to avoid interparticle bridging, and a large excess of PEs coming from each adsorption step that has to be removed prior to the generation of the next layer. To circumvent these limitations, alternative formulations were proposed. This is the case for the Precipitation-Redispersion pathway initially developed to coat bare nanoceria NPs ($CeO_2$) with a single layer of ion-containing polymers [16] and used later on iron oxide ($\gamma$-$Fe_2O_3$) NPs [17-19]. Here, when a NP dispersion was mixed to a PAA solution at acidic pH, the solution underwent an instantaneous and macroscopic precipitation. The phase separation resulted from the multisite adsorptions *via* H-bonding of the uncharged acrylic acid moieties on the particle surfaces. As the pH of the precipitate was further increased by addition of a base, the NPs redispersed spontaneously, yielding a clear and concentrated solution of individually PAA coated particles. With the Precipitation-Redispersion pathway, large amounts of NPs (~ 1 g of oxide) can be prepared at once, and can serve as a basis for further functionalization .

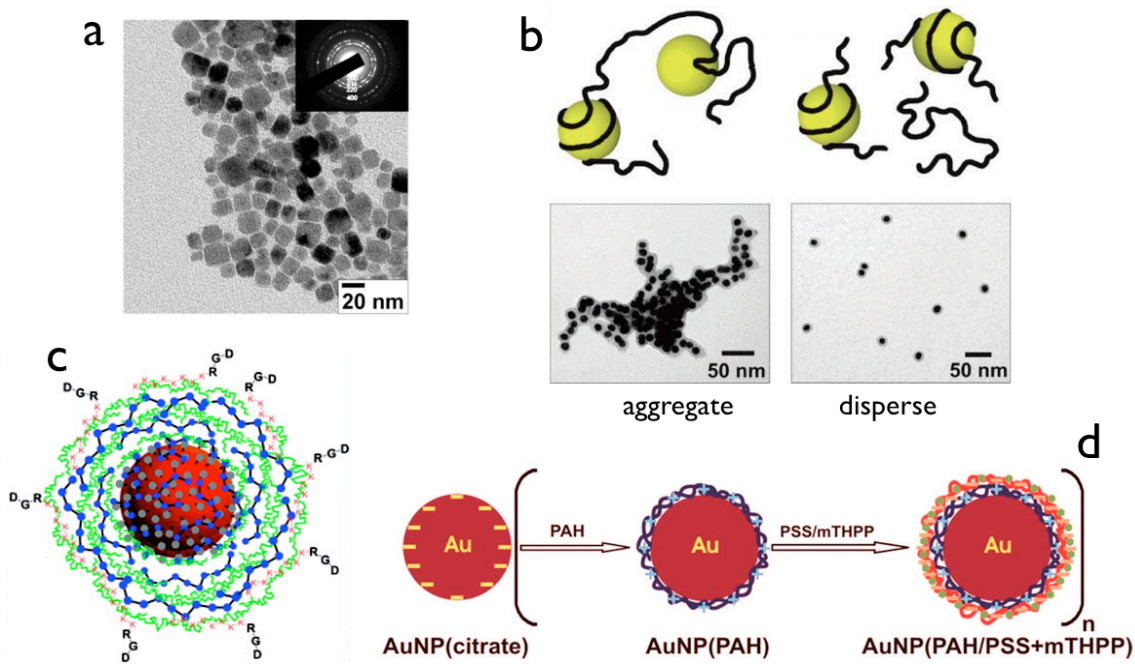

*Figure 1 :* Construction and organization of discrete hybrid nano-objects resulting from the electrostatic complexation of inorganic nanoparticles with polyelectrolytes. **a)** Transmission electron microscopy image of 18 nm $Cu_2O$ nanocubes prepared in the presence of poly(sodium acrylate) of molecular weight 1200 g $mol^{-1}$ [6]. **b)** Upper panel : representations of particles being aggregated by bridging flocculation (left) and being individually wrapped by polymer chains (right); Lower panels : electron micrographs of aggregated gold particles (left) and of particles predominantly unsheathed individually (right) [5]. **c)** Illustration of a LbL self-assembly strategy for multifunctional silica NPs containing luminescent and magnetic probes [11]. **d)** Principle of loading a porphyrin-based anticancer drug (mTHPP) at the surface of gold NPs using standard ELbL method [12].

## 2.3 - Towards multilayers

After the first layer, PE multilayers were deposited on gold NPs [12,20**] and nanorods [4,15*], as well as on magnetic and multimodal objects [11,21] to design multifunctional nano-objects for drug delivery, imaging and photothermal therapy. The adsorbed PEs were classically those used for flat surface coverage, e.g. PAH, PSS and PDADMAC. With the L*b*L deposition, NPs could be made anionic or cationic with up to 10 layers [5,12] and their interactions with living cells [4,20**] or extracellular matrix [15*] accurately monitored. Wilson *et al.* showed for instance that gold nanorods modified by PE multilayers minimized cytotoxicity and that polyanion-terminated rods interacted more strongly with type I collagen, the constituent of the extracellular matrix [15*]. This was not the case for the particles with a polycation in the last layer. In addition to the enhanced stability in physiological media, multilayered NPs were also synthesized to encapsulate actives and drugs [12,20**]. Schneider *et al.* designed mutifunctional gold particles where doxorubicin, a very potent deoxyribonucleic acid (DNA) intercalation agent and anti-cancer drug could be incorporated and in the same time released *in vitro* [20**]. More recently, Reum *et al.* intercalated with this technique up to three layers (in addition to those made of PEs) of a water insoluble porphyrin–based anti-cancer drug inside the ultrathin coating multilayers, increasing hence the drug load efficiency considerably [12]. Schematic representations of multilayered structures are sketched in Fig. 1c [11] and Fig. 1d [12].

### 2.4 - Block copolymers

Adsorbing ion-containing copolymers, instead of homopolymers as seen previously, is a technique that aroused a strong interest during the last years. With copolymers, the formulation resembled that described for the adsorption of the first PE layer (Section 2.2) [19]. Polyvalent hyperbranched PEs were used to tune the thermal and pH responses of gold NPs and reversibly induce aggregation [22]. Copolymers with a poly(ethylene glycol) (PEG) neutral block were also exploited together with iron [23] and titanium [24] oxides and yielded an enhanced stability in physiological media thanks to the neutral corona surrounding the particles. With charged-neutral block copolymers, it was also found that the *direct mixing* may result in the formation of aggregates or clusters of NPs, with sizes in the range of 20 nm to 200 nm [25,26]. These latter reports stressed the needs of building NP clusters with controlled sizes and shapes for various applications, with the help of macromolecules acting as binders.

# 3 – Clustering and bulk complexation

For applications, particulates with sizes in the 100 nm – 1000 nm range may be required but yet remain difficult to generate by soft chemistry. This is the case for gold, iron oxide NPs and quantum dots. Several issues become critical when the particle size increases, including colloidal stability, sedimentation and loss of physical properties. In order to circumvent these limitations, co-assembly strategies involving sub-10 nm NPs were elaborated [16,17,27*-29**]. NPs were arranged into clusters of different dimensions and morphologies, with aggregation numbers ranging from few units to several thousands. Within the clusters, the NPs are at close contact and exhibit a structural disorder of their centre-of-masses (like in a glass), with volume fraction reaching 20 to 30 vol. %. For the fabrication of stable NP clusters, the outline will follow that of single NPs.

### 3.1 – PE-assisted synthesis of nanoparticle clusters

As shown before, PEs introduced in the reduction process of metallic ions can serve as a nanoreactor or template for the synthesis of individual NPs. For a different outcome, PEs are also used to build highly stable aggregates or clusters of small primary particles. These clusters are said to be hierarchical because they involve different length scales, such as the size of the single NPs (~ 10 nm) and that of the overall aggregate. Yin and coworkers proposed a synthesis of $Fe_3O_4$ magnetic clusters based on the hydrolysis of Fe(III) in diethylene glycol in the presence of PAA [29**,30]. PAA ($M_W$ = 1800 g mol$^{-1}$) was selected by these authors because of the strong coordination of carboxylate groups with iron cations on the magnetite surface. With this synthesis, 30 to 180 nm clusters were produced. As a result of the balance between magnetic attraction and electrostatic repulsion between clusters, the dispersion formed a colloid crystal that diffracted light at wave-length that can be tuned from red to blue by increasing the magnetic field (Fig. 2a). The capability of fast on/off switching of the diffraction by magnetic fields makes these magnetochromatic microspheres suitable for applications such as color display, rewritable signage, and sensors [30]. In the same spirit, Adschiri *et al.* carried out a one-shot synthesis of palladium clusters, coined ''mosslike hybrid particles'' by the authors, using the reduction of Pd salts by ascorbic acid in the presence of DNA [27*] or of PAA [31]. This reduction approach yielded spherical Pd-clusters of sizes between 30 and 100 nm, the primary NPs having a diameter of 5 nm. The mosslike hybrid materials were thereafter deposited and dried on a silica chip, resulting in a highly sensitive hydrogen sensor device directly applicable to fuel cell technology [27*,31].

## 3.2 – Direct mixing

At charge stoichiometry, the *direct mixing* pathway using oppositely charged NPs and PEs leads to the clustering and eventually the flocculation of the NPs. The clustering of NPs revealed however interesting features, in particular in the context of DNA compaction [32*]. Zinchenko and coworkers investigated the compaction of single DNA molecules (57 μm contour length, 166000 base pairs) by cationic silica NPs to reproduce a model of the histone-core particles, which are the elementary bricks of chromatin. Using fluorescence microscopy, these authors found that with increasing NP concentration (the DNA concentration being fixed at 100 ppm), the compaction of a single chain increased and displayed well-defined intermediate states including the chromatin-like bead-on-a-string structures. These structures consisted of NP connected by a thin DNA thread. However, contrary to the periodic structure of natural chromatin, NPs were distributed in a non periodic way as shown in Fig. 2b, and could even form aggregates along the chain [32*]. Other clustering strategies allowing interchain associations were also attempted using magnetic NPs for the evaluation of bimodal magnetic-fluorescent contrast agents [28].

## 3.3 – Formulation approach using intermediates
Interesting colloids can be generated from three strongly interacting components instead of two as before. These components comprised all inorganic NPs and a pair of organic moieties, which can be alternatively a multivalent counterion [33], a PE [34] or a microsphere [18]. By playing on the different concentration and charge ratios, Schneider and Decher have formulated colloidal dispersions made of fairly monodisperse nano- and micro-"pouches" containing a targeted load of gold or of iron oxide NPs [33]. The co-assembly occurred spontaneously by electrostatic complexation, and provided spherical soft colloids, as illustrated in Fig. 2c. These hybrid nano/micro-"pouches" are potentially interesting for therapeutic applications. In the presence of salt at the isotonic level of physiological media (9 g L$^{-1}$), the nano-"pouches" dissolved and their NP load was entirely released. With a pair of oppositely charged PEs, namely poly(aspartic acid) and chitosan, Wang and coworkers realized similar composite microspheres, incorporating either magnetite particles or quantum dots [34]. Using again a 3-component complexation scheme, Fresnais *et al.* developed an easy and versatile route for the fabrication of magneto-fluorescent nanowires [18]. Using the *desalting transition* pathway (described in Sect. 3.5), 1 - 50 μm long nanostructured aggregates comprising 7 nm magnetic particles and 180 nm fluorescent microspheres (with PEs as "gluing" agents) were generated. The hybrid architectures maintained the magnetic and optical properties of the initial constituents, as evidenced by fluorescence microscopy and by the in-phase response of the wires to an external rotating magnetic field.

### 3.4 – Diblock copolymers and formulation issues

To slow down the kinetics of association observed at charge stoichiometry, complexation schemes using neutral-charged block copolymers (instead of homoPEs) were also studied [25,26]. The primary goal was to assess the analogy between core-shell electrostatic-based clusters and polymeric micelles made from amphiphilic molecules, as suggested by Cohen-Stuart and coworkers [35,36]. According to these authors, a micelle can be regarded as a droplet of a new phase that has been arrested in its growth because part of its constituents does not participate in the phase separation. This strategy appeared successful : by changing the molecular weights of the charged and neutral blocks, NP aggregates of controlled sizes and morphologies could be obtained [25]. From these studies, it became clear however that under strong driving forces like electrostatic interactions, the final products also depended on the preparation and mixing conditions, such as the pH, concentration, mixing speed. Chapel and coworkers investigated the formation of hybrid complexes composed of $CeO_2$ NPs and charged-neutral diblock copolymers by tuning the mixing [37**] and the interaction pathways [38**]. It was found that the key factor controlling the polydispersity and the final size of the complexes was the competition between the reaction time of the components and the homogenization time of the mixed solution. This was achieved by comparing various formulation techniques such as high speed injection, screening of interactions and microfluidic mixing. Interestingly, these authors pointed out that a process-dependent formulation seen as a drawback can be turned into an advantage: different properties can be developed from different morphologies while keeping the chemistry constant [37**,38**].

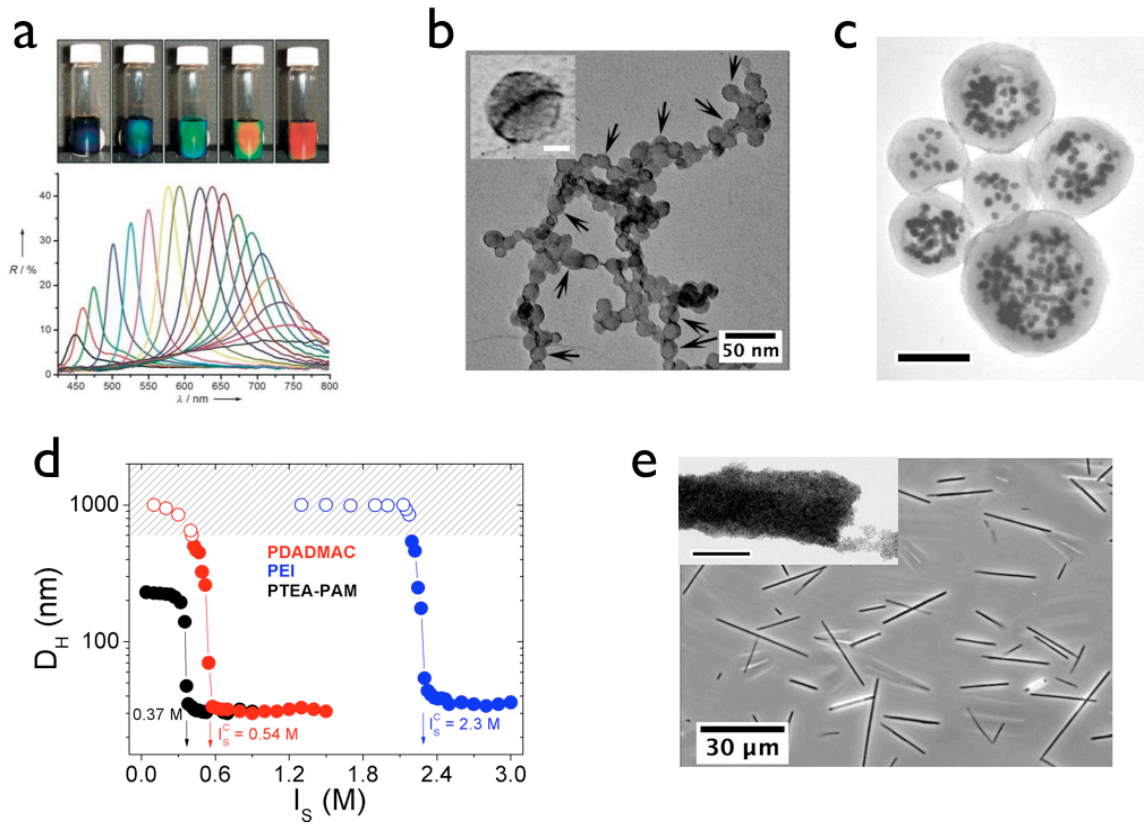

***Figure 2 :*** *Construction and organization of hierarchical hybrid nano-/micro-objects resulting from the electrostatic complexation of inorganic nanoparticles with polyelectrolytes. **a)** Upper panel : photographs of colloidal crystals of NPs clusters formed in response to an external magnetic field ; lower panel : dependence of the Bragg reflection spectra on the distance of the sample from the magnet [29\*\*].**b)** Transmission electron microscopy images of the local arrangement of a DNA chain on 15 nm silica NPs. The arrows and inset indicate NPs onto which DNA wrapping is clearly visible [32\*].**c)** Transmission electron micrographs of individual micro-"pouches" formed through the flocculation of a three component system, PAH, trisodium citrate and gold NPs. The bar is 100 nm [33]. **d)** Desalting Transition observed for oppositely charged iron oxide NPs and PEs and monitored by dynamic light scattering. $D_H$ denotes the hydrodynamic diameter and $I_S$ the ionic strength of the dispersion [19].**e)** Optical microscopy images of 15 µm magnetic nanowires obtained by electrostatic co-assembly of 7 nm iron oxide NPs and oppositely charged PEs [17\*\*,18].*

## 3.5 – Desalting transition

The protocols for mixing oppositely charged species based on the *desalting transition* were drawn from molecular biology and were developed for the *in vitro* reconstitutions of chromatin [32\*]. In the context of inorganic NPs, the *desalting transition* was studied by Berret and coworkers in a series of papers dealing with the controlled clustering of $CeO_2$ and $\gamma\text{-}Fe_2O_3$ sub-10 nm particles. The protocol consists in two steps, first the screening of the electrostatic interactions by bringing the dispersions of oppositely charged species to high salt concentration, and second in removing the salt progressively by *dialysis* or by *dilution*. With this technique, the oppositely charged species are intimately mixed in solution and do not interact owing to the electrostatic screening. As suggested by Stoll and Chodanowski using Monte Carlo simulations (see [39] and references therein), an abrupt transition between a disperse and an aggregated state of particles was found at a critical ionic strength that depends on the nature of the PEs [19]. The desalting kinetics was first investigated using charged-neutral block copolymers, and later on generalized to homoPEs, such as PEI and PDADMAC. An illustration of the transition is shown in Fig. 2d. One of the most interesting results obtained via the *desalting transition* was the formation of highly rigid magnetic nanowires with length comprised between 1 and 100 μm [17**]. The nanowires were synthesized by dialysing γ-$Fe_2O_3$ and PEs in the presence of a constant magnetic field of 0.1 Tesla. Fig. 2e shows images of the nanowires obtained by transmission electron microscopy (inset) and by optical microscopy. These nanowires are devoted to applications in nanomedicine, where they can be used as nanodevices (tips, tweezers and actuators) at the scale of the cells.

# 4 - Functional thin films, colloids or capsules through Electrostatic Layer *by* Layer assembly

A very smart use of the electrostatic interaction between oppositely charged building blocks was pioneered by Iler [40] in the sixties to generate all-nanoparticles multilayers, then rediscovered and brought to fame in the nineties by Decher [13] using PE complexes. Nowadays, the layer-by-layer (L*b*L) sequential adsorption method is an efficient tool to elaborate multiscale organic/inorganic tunable functional coatings, colloids or capsules. With the advent during the last decade of controlled and reproducible inorganic NPs synthesis pathways, this layering approach which takes now benefit from both the inorganic and organic world has literally exploded. This blooming area where new developments are continuously put forward is getting more mature with years but some challenges still remains.

## 4.1 - Thin films

Uses of an electrostatic L*b*L approach (EL*b*L) to generate hybrid functional films will be briefly discussed through the prism of some interesting developments pointing out advantages, drawbacks and limitations of the process. The EL*b*L hybrid multilayers present in some specific cases a very good stability and reusability making this assembly process a very good candidate for industrial scale-up over other methods including spin or drop-casting techniques for example. Priya *et al.* [41] have generated $TiO_2$/PE multilayer thin films in order to photodegrade rhodamine B, a widespread organic pollutants used in many industrial processes like paper dyeing or dye laser production. These authors have shown that the robustness of the EL*b*L coating enables the $TiO_2$ catalyst to be reused several times with the same efficiency, opening then the possibility of direct applications in the textile industry waste water treatment, an acute environmental problem.

On a different aspect, EL*b*L is generally performed from environmentally friendly waterborne solutions. In some sensitive areas like film electronics however, aqueous solutions can generate specific problems. Nakashima *et al.* [42] have developed a PEs/carbon nanotube EL*b*L multilayer from ionic liquids which prevented any water corrosion and hydration processes to occur. These authors have also shown that conductivity performances of transparent films made either from ionic liquids or water solutions were equivalent, suggesting that those ionic liquids, considered as green ''processing solvents'', were indeed a very good alternative to avoid corrosive attacks or hydration reactions.

In term of process, the electrostatic interaction can be advantageously supplemented by other type of forces as shown by Dey *et al.* [43] where oppositely charged ferromagnetic magnetite ($Fe_3O_4$) NPs were grown alternatively under the presence of a magnetic field (0.3 Tesla). This *field-assisted* EL*b*L resulted in a well oriented NPs magnetic domains perpendicular to the substrate within a denser/compact multilayer with fewer defects than with the sole electrostatic driven interaction opening an interesting route for the generation of magnetically responsive functional coatings. In the same vein, the electrostatic and hydrogen bonding interactions can also be ingeniously associated to built a photoluminescent pH-responsive hydrogel, as shown recently by Kharlampieva *et al.* [44**]. The key idea here was the formation of an alternate assembly of classical EL*b*L strata confining tightly quantum dots (PSS/PAH + anionic CdTe) and H-bonded layers (poly(methacrylic acid)/poly(vinyl pyrrolidone)) subsequently cross-linked to form a gel stratum bringing the pH responsiveness (Fig. 3a). This approach allowed tailoring the layers architecture by a simple manipulation of the thickness of both the EL*b*L and gel layers, offering then a robust hybrid multilayer device with *reversible* optical responses.

EL*b*L has obvious drawbacks that sometimes may turn into real advantages. The electrostatic deposition of PE layers is mainly controlled by their phase behavior in the bulk solution resulting in some specific cases to the erosion of the multilayer stack rather than its further growth [45]. Magdassi *et al.* have likely experienced such drawback and turned it into an advantage to trigger the sintering of silver NPs at room temperature generating very efficient conductive plastic films *via* inject printing [46*]. PAA stabilized Ag NPs were indeed deposited onto a solid substrate forming a 0.5 μm thick array. The substrate was then dipped into a strong polycation (PDADMAC) solution resulting in the coalescence of the NPs *via* charge neutralization and a huge increase of the film conductivity (70 μΩ.cm). The key idea here was that for a specific concentration (PDADMAC/Ag ratio 10 times lower than in the bulk) the presence of the polycations in the bulk solution likely *desorbs or ''erodes''* the PAA ligands layer inducing the coalescence of the unprotected Ag NPs; A very interesting concept in the field of plastics electronics and beyond. Illustrations of such films are provided in Fig. 3b.

Even though numerous developments are appearing every day, full control over all important parameters of this layering process is still lacking. A good illustration of some of the limitations is given by the work of Mc Clure *et al.* [47*] in the sprawling domain of photovoltaic. These authors have used the EL*b*L approach to generate a robust heterojunction with a high interfacial area between oppositely charged inorganic nanosized semi-conductor and conjugated polymers. A series of thin film devices were fabricated on indium tin oxide electrodes using cationic CdSe nanorods layered with anionic conjugated polythiophene-based photoactive polymers. Those films showed somehow limited photovoltaic performance due to low short circuit current even though charge transfer was evidenced in the polythiophene/CdSe nanocomposite thin films. A thorough analysis pointed out several parameters of the process affecting the carrier transport and not necessarily encountered in less sensitive applications. Indeed, the roughness of the generated thin films, the nanorod-nanorod contact and orientation and the nature of the surface ligands are clearly general process dependent parameters that should be improved or better controlled in the future to get superior performances in any domain.

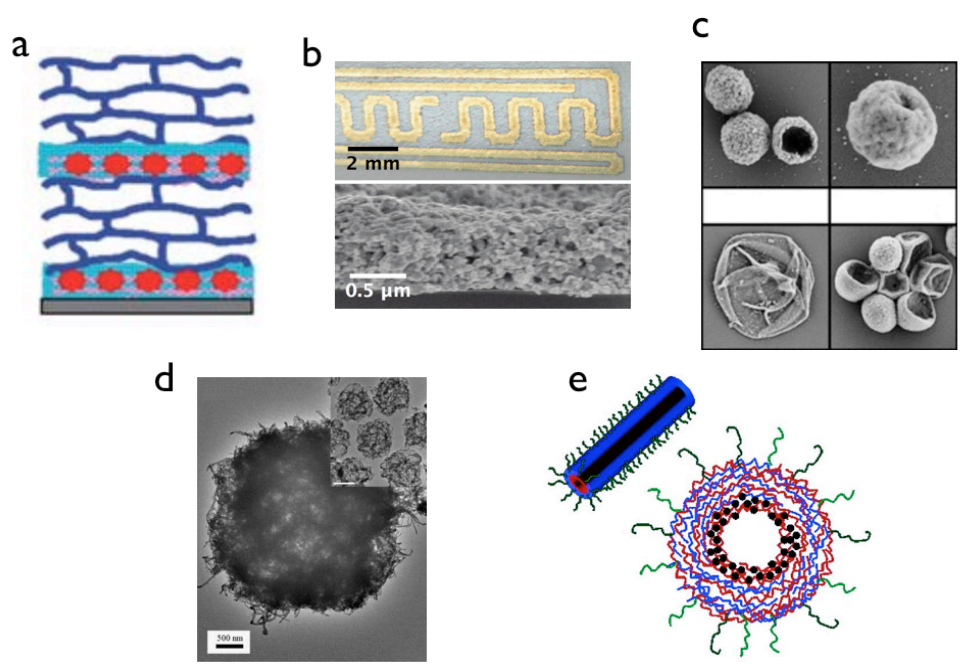

*Figure 3 : Typical examples of thin films and capsules generated from NPs and PEs building blocks via the Electrostatic Layer-by-Layer complexing route. a) Sketch of an optically active hybrid hydrogel made through alternating assembly of electrostatically-bound and hydrogen-bonded layers loaded with quantum dots [44\*\*]. b) Triggered coalescence of conductive silver NP films. Upper panel: macroscopic image of a pattern printed on Epson photo paper; lower panel : High resolution SEM images of the surface [46\*]. c) SEM images of thermally shrunk capsules assembled from gold NPs and PEs at different temperatures (70 and 80 °C from left to right) and different NPs surface coverage (7% and 28 % from top to bottom) [48]. d) Electron micrographs of multiwalled CNT capsules after removal of the PEs scaffold and PMMA sacrificial template via calcination. Inset: lower magnification (scale bar is 2 µm) [49]. e) Schematic axial cross section of a heterostructured magnetic hollow nanotube made from the assembly of $Fe_3O_4$ NPs and PAH after dissolution of the sacrificial polycarbonate membrane [50].*

## 4.2 - Nanoparticles and microcapsules

Beyond macroscopic functional surfaces and thin films and their great scientific and technological potential, EL*b*L serves as a versatile tool box to engineer functional micro-sized hollow capsules from sacrificial colloidal particles. This approach originally pioneered by Möhwald and colleagues [48,51,52] gives a very precise control over the size, shape, chemical composition, wall thickness and functionality of the engineered capsules with particular added-value in the biosensors and drug delivery domains [53]. As in the thin films area, the presence of inorganic NPs bearing intrinsic properties not seen with macromolecules is definitely boosting this domain. Examples of such capsules seen by scanning electron microscopy (SEM) are illustrated in Fig. 3c [48].

Very recently, to catch up with the exponentially growing carbon nanotube (CNT) literature, microcapsules made out from single [54] and multiwalled [49] CNT were also elaborated. In both cases, a colloidal particle template (PMMA or silica) was sequentially layered with CNTs and oppositely charged synthetic PDADMAC or functionalized beta-1, 3-glucans polysaccharides (Fig. 3d). The template was then removed either *via* calcination or HF dissolution generating hollow CNT capsules with uniform sizes and tunable wall thicknesses with potentially new applications such as in catalysis, biomedical devices or lithium ion batteries.

At a smaller scale, NPs can also be engineered with a sophisticated design *via* EL*b*L as shown recently by Schneider *et al.* [20**] on a gold-based core/shell drug delivery system consisting of cytotoxic stealth carrier particles (see Sect. 2.2 and 2.3). The excellent stability and size control of such modular nanocarriers make them good candidates for enhanced permeability and retention targeting.

Beyond sphericity, the EL*b*L can also be implemented to generate multifunctional nanorods and tubes. Wu et *al.* developed a novel synthesis pathway to broaden the applications of the wide band gap ZnO semiconductors [55]. Noble metals, metal oxides, or metal sulfides were assembled at room temperature onto the ZnO template to generate nanorod-based hybrid materials including $ZnO/CeO_2$, $ZnO/CdS$, and $ZnO/Ag$. PAH/PSS PEs were first layered onto the surface of the rods creating a negative environment suitable for the subsequent adsorption of positively charged metals ions ($Ce^{3+}$, $Cd^{2+}$, and $Ag^+$), the key point of the overall process. They were then either reduced by sodium borohydride such as $Ce^{3+}$ and $Ag^+$ or reacted with $S^{2-}$ such as $Cd^{2+}$ to make a uniform NPs coating. This approach can efficiently be extended to others high-quality 1D hybrid nanomaterials. In the same line, magnetic colloidal capsules with high aspect ratio have generated a large body of research recently because they can be manipulated by means of magnetic fields and gradients, and the inner part can be loaded with small molecular weight actives released on demand later on. Lee *et al.* have developed an approach to generate heterostructured magnetic submicronic nanotubes *via* EL*b*L in the pores of track-etched polycarbonate membranes [50]. The membrane was then dissolved leaving individual superparamagnetic tubes made out from different PEs/magnetite ($Fe_3O_4$) NPs assembly. A schematic view of the tube axial cross section is shown in Fig. 3e. pH triggered release of anionic model molecules previously loaded into the wall multilayer was monitored. Those magnetic carriers can then be used as drug delivery agents driven to target sites *via* a magnetic field, *in vivo* or *in vitro*.

Last but not least, controlling the size and distribution of NPs on top of PE multilayer capsules is generally of paramount importance. Using a straightforward admixing approach, Parakhonskiy *et al*. have recently shown that in the low concentration regime gold NPs can be assembled onto PE capsules with a non aggregated distribution in sharp contrast with the concentrated regime leading to aggregates [56]. Such method of controlled NP adsorption can be very helpful for the targeted delivery and release of drugs [51].

# 5 - Direct applications

Throughout this review we have clearly seen that the complexation between polymers and NPs is offering a good platform for engineering versatile hybrid structures. The emergence of novel building-blocks and processing at the nano scale has generated numerous applications in various high-tech sectors like electronics, energy, photonics or biomedicine but also in more classical domains as automotive, textile or paper industry.

## 5.1 - Low-tech applications

NPs/macromolecules electrostatic complex derivatives do not necessarily show up only in high-tech sectors. Long standing and mature industries can also benefit from their synergies. Peng *et al.* developed conductive paper from EL*b*L assembly from indium tin oxide NPs and PEs (PEI, PSS) onto wood fibers manufactured using traditional paper industry processes [57]. Values of electric conductivity 6 orders of magnitude higher than that of plain paper were typically measured ($5.2 \times 10^{-6}$ S cm$^{-1}$ for in-plane, $1.9 \times 10^{-8}$ S cm$^{-1}$ through-the-thickness), making possible the engineering of electronic devices such as capacitor or transistors from a cost-effective cellulose pulp.

Metal corrosive attacks are one of the big challenges of the automotive and air transportation industries. Grigoriev *et al.* have developed mesoporous negatively charged silica micro-containers filled with benzotriazole inhibitor *via* EL*b*L assembly ($SiO_2$/PEI/PSS(BTA/PSS)$_n$) and embedded into sol-gel coating ($SiO_2$–$SiO_x$:$ZrO_x$) for the protection of aluminum alloys [58]. Upon corrosive attack the local pH change influences the PSS/benzotriazole shell, inducing its dissolution with the inhibitor release and subsequent healing of the corrosion site. This self-healing anti-corrosion coating presents then a very interesting active feedback loop combined with a sustained release of inhibitor exactly at the corrosion-damaged sites.

In the textile industry, durability of superhydrophobic cotton fabrics against washing remains a great challenge. Indeed, a fast drop of the static contact angle with water is usually observed after only 10 washing cycles rendering the fabric somehow useless. To circumvent such drawbacks, Zhao *et al.* [59] have elaborated a robust process in which highly hydrophilic cotton fibers are turned superhydrophobic *via* the electrostatic wrapping of silica NPs and oppositely charged PEs (cotton/PAA/ (PAH/SiO2)$_n$). With at least 5 layers, slippery superhydrophobicity with a small hysteresis was achieved giving rise to remarkable buoyancy and an enhanced durability up to 30 washing cycles.

## 5.2 - High-tech applications

High-techs applications can fall into material science and biomedical domain as reflected by their respective literature weight.

*5.2.1 - Material science domain*

In the very crucial domain of alternative sources of energy, Wang *et al.* [60] have demonstrated the pertinent use of EL*b*L assembly to create hybrid bipolar plate coatings composed of both electrically conductive graphite platelets and silica NPs to maximize proton exchange membrane fuel cell efficiency *via* an enhanced surface wettability and a reduced contact resistance. In the same area, Taylor *et al.* [61] have generated ultrathin layers made out of single-walled CNTs functionalized Pt catalyst layered with the help of some specific PEs (Nafion, PEI, poly(aniline)) with excellent homogeneity characteristics yielding a power density 4-fold higher than membrane electrode assemblies produced using conventional methods.

Due to their very small size, semiconducting NPs exhibits attractive electrical and optical properties compared to their bulk counterpart. Chan *et al.* have shown recently interesting electrical bistability when these NPs are embedded in polymers matrices, a key feature for microelectronics applications [62]. Pal's group [63*] has elaborated all-NPs 2D arrays of CdSe *via* EL*b*L assembly from oppositely charged NPs with tunable size and shown a reversible size-dependent bistability down to the quantum dot regime opening pathways to engineering CdSe quantum dots as ROM & RAM memory elements with a density of one bit per particle [64]. Photonics applications also profited from the generation of newly shaped materials [29**,30] capable of diffracting light continuously through the application of an external magnetic field, see Sect. 3.1.

*5.2.2 - Biomedical domain*

Biomedical applications are certainly one of the high-tech sectors which benefit the most from the NP/macromolecules complexation synergies.

The need for real-time monitoring of physiological key functions has indeed triggered the fast blossoming of the biosensing field for which EL*b*L derivative films or capsules are ideal candidates [53]. One example of particular interest is the monitoring of glucose concentration in blood to diagnose, cure or prevent diabetes mellitus, a metabolic disease affecting more than hundred million people in the world. Different strategy have been put forward recently using nanocomposite biosensors made out from NPs and/or oppositely charged functional macromolecules [65,66]. Li *et al.* [67] have developed an optically active EL*b*L multilayer film to promote rapid high-throughput detection when integrated into a lab-on-chip device instead of a colloidal-based sensor [68]. The multilayer composed of CdTe quantum dots, oppositely charged PEs and negatively charged active glucose oxydase enzyme (GOx) generated hydrogen peroxide ($H_2O_2$) when contacted with a glucose solution. This in turn generated quantum dots surface defects leading to the quenching of their luminescence as a function of the glucose concentration in the blood serum.

In the same vein, NPs based electrochemical [69,70] or optical biosensors [71] have attracted much attention as a potentially simple diagnostic tool as shown elegantly by Xia *et al.* [72*]. The colorimetric sensing strategy developed by these authors was based on the finding that the presence of a positively charged, water-soluble, conjugated PE inhibited the ability of single-stranded DNA to stabilize bare gold NPs, leading to the solution aggregation. This feature allowed therefore a readily detectable change in their color, whereas no color change was observed with double-stranded or "folded" DNA structures. The electrostatic interaction indeed regulates the interaction between the conjugated polymer and various DNA forms of different hydrophobicity. This differential affinity enables the development of sensitive colorimetric assays where convenient-picomolar concentrations of target DNA are readily detected with the naked eye.

Developing efficient molecular imaging assays for studying biological systems *in vivo*, is of particular importance nowadays. Kim *et al.* [11] have developed functional NPs as magnetic resonance imaging (MRI) contrast agents [73] that can target specific diseased cells using a simple non covalent functionalization strategy. From a silica NP containing a luminescent $[Ru(bpy)_3]Cl_2$ core and a anionic paramagnetic shell EL*b*L assembly of negative PSS and active cationic gadoterate meglumine (Gd-DOTA) oligomers was generated. The very high relaxivities measured were due to the highly disordered and hydrophilic nature of the multilayer which allows ready accessibility of water molecules to the Gd-centers for efficient water proton relaxation. Those core-shell NPs were subsequently functionalized with positively charged targeting peptides to lead to cancer-specific multifunctional NPs for optical and MRI of HT-29 human colon cancer cells. This general approach can be easily engineered to target different other pathologies.

Beyond imaging, the targeted and triggered release of active load is also of paramount importance in biomedicine and generates nowadays a very active scientific and technological field. Lee *et al.* [74] have developed for example pluronic/PEI crosslinked nanocapsules with embedded magnetite nanocrystals for magnetically triggered delivery of small interfering ribonucleic acid (siRNA) and its great therapeutic potential. The positively charged capsules were then covered with negatively charged siRNA-PEG conjugates. Those stable PE complexes were more efficiently taken up by cancer cells in response to an external magnetic field enhancing intracellular uptake and effectively suppressing the green fluorescent protein expression in cancer cells without showing any severe cytotoxicity. This approach constitutes likely an interesting physico-chemical platform for magnetically triggered delivery of various negatively charged therapeutic agents as well as MRI diagnosis tool box.

# 6 - Conclusion

This review featured the great potentials the electrostatic complexation between NPs and PEs is offering in term of sound science and exciting applications. From a literature survey, we have identified three major formulation processes: the *direct mixing* route, the *desalting transition* pathway and the universal *Layer-by-Layer* method. Those different complexation routes gave birth to four main active areas of research with distinct features but permeable boundaries: the synthesis of stable NPs , the one- or multilayer coating of single NP's, the generation of hybrid colloidal capsules or thin films and the bulk clustering. With specific examples, we have illustrated their current developments, pointed out some of their drawbacks and highlighted some of their future applications in materials science and nanomedicine domains.

# Acknowledgements

We thank Jean-Christophe Castaing, Andrejs Cebers, Jérôme Fresnais, Eléna Ishow, Sébastien Lecommandoux, Patrick Maestro, Nathalie Mignet, Mikel Morvan, Julian Oberdisse, Régine Perzynski, Ling Qi, Olivier Sandre, Christophe Schatz, Sribharani Sekar, Minhao Yan for fruitful discussions. This research was supported in part by Rhodia (France), by the Agence Nationale de la Recherche under the contracts BLAN07-3_206866 and ANR-09-NANO-P200-36, by the European Community through the project : "NANO3T—Biofunctionalized Metal and Magnetic NPs for Targeted Tumor Therapy", project number 214137 (FP7-NMP-2007-SMALL-1) and by the Région Ile-de-France in the DIM framework related to Health, Environnement and Toxicology (SEnT).

**TOC**

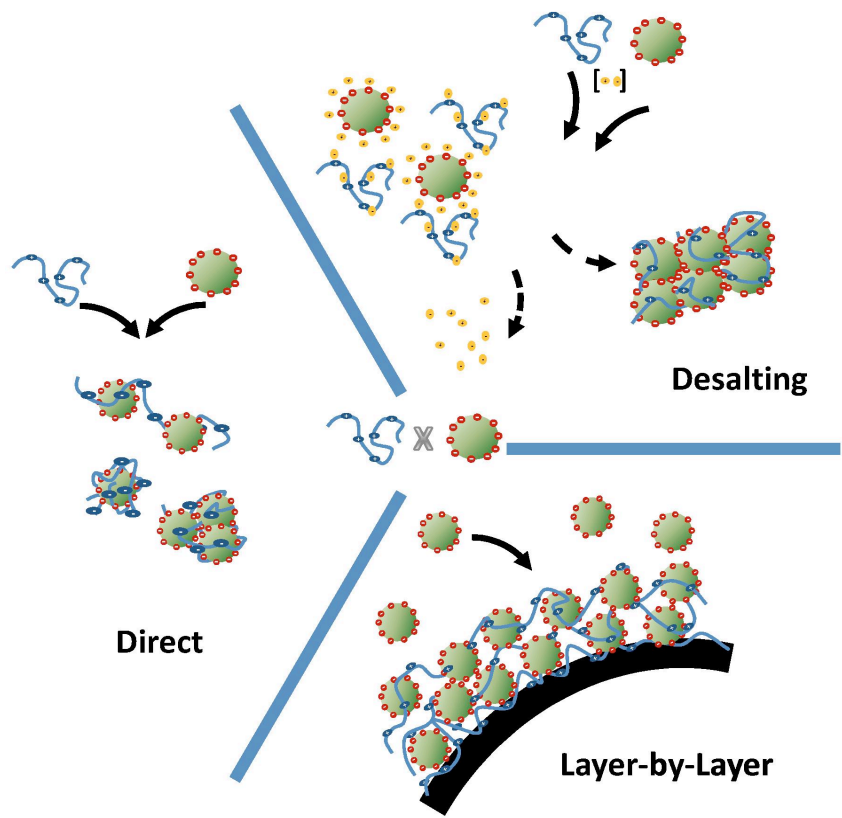